\documentclass{aa} 
\usepackage{graphics,astron}

\titlerunning{ISO observations of the Wolf-Rayet galaxy \object{NGC 7714}}
\title{ISO observations of the Wolf-Rayet galaxy \object{NGC 7714} and its
companion \object{NGC 7715}}

\authorrunning{B.O'Halloran et~al}
\author{B.O'Halloran\inst{1}, L. Metcalfe\inst{2}, M. Delaney\inst{1},
 B. McBreen\inst{1}, R. Laureijs\inst{2}, K. Leech\inst{2}, 
 D. Watson\inst{1} \and L. Hanlon\inst{1}}

\offprints{halloran@bermuda.ucd.ie}

\institute{Physics Department, University College, Belfield, Dublin 4, Ireland
   \and ISO Data Centre, Astrophysics Division, Space Science Department of ESA,
   Villafranca, P.O.  Box 50727, E-28080 Madrid, Spain} 
   
\date{Received 14 April 2000 / Accepted 7 June 2000} 
   
\thesaurus{03(11.09.1 NGC 7714; 11.09.1 NGC 7715; 11.09.2; 11.19.3)} 

\begin{document}
\maketitle

\begin{abstract}

The interacting system \object{Arp 284} consisting of the Wolf-Rayet galaxy
\object{NGC 7714} and its irregular companion \object{NGC 7715} was observed
using the Infrared Space Observatory.  Deconvolved ISOCAM maps of the galaxies
using the 14.3 $\mu$m, 7.7 $\mu$m and 15 $\mu$m LW3, LW6 and LW9 filters, along
with ISOPHOT spectrometry of the nuclear region of \object{NGC 7714} were
obtained and are presented.  Strong ISOCAM emission was detected from the
central source in \object{NGC 7714}, along with strong PAH features, the
emission line [Ar II], molecular hydrogen at 9.66 $\mu$m and a blend of features
including [S IV] at 10.6 $\mu$m.  IR emission was not detected from the
companion galaxy \object{NGC 7715}, the bridge linking the two galaxies or from
the partial stellar ring in \object{NGC 7714} where emission ceases abruptly at
the interface between the disk and the ring.

The morphology of the system can be well described by an off-centre collision
between the two galaxies.  The LW3/LW2, where the LW2 flux was synthesized from
the PHOT-SL spectrum, LW9/LW6 and LW3/LW6 ratios suggest that the central burst
within \object{NGC 7714} is moving towards the post-starburst phase, in
agreement with the age of the burst.  Diagnostic tools including the ratio of
the integrated PAH luminosity to the 40 to 120 $\mu$m infrared luminosity and
the far-infrared colours reveal that despite the high surface brightness of the
nucleus, the properties of \object{NGC 7714} can be explained in terms of a
starburst and do not require the prescence an AGN.

 \end{abstract}

\section{Introduction}

  Wolf-Rayet galaxies are a subset of emission line and H II galaxies whose
integrated spectra have a broad emission feature around 4650\AA~which has been
attributed to Wolf-Rayet stars.  The emission feature usually consists of a
blend of lines namely, HeI $\lambda$4686, CIII/CIV $\lambda$4650 and NIII
$\lambda$4640.  The CIV $\lambda$5808 line can also be an important signature of
Wolf-Rayet activity.  Wolf-Rayet galaxies are important in understanding massive
star formation and starburst evolution
\cite{iso-schaerer-1998aj,iso-mashesse-1999aa}.  The Wolf-Rayet phase in massive
stars is short lived and hence gives the possibility of studying an
approximately coeval sample of starburst galaxies
\cite{iso-metcalfe-1996aa,iso-rigopoulou-1999esa}.  The first catalog of
Wolf-Rayet galaxies was compiled by Conti (1991) and contains 37 galaxies.  A
large number of additional Wolf-Rayet galaxies have been identified and a new
catalog containing 139 galaxies has been compiled by Schaerer et al. (1999).

\begin{figure*} \resizebox{\textwidth}{!}
{\includegraphics*{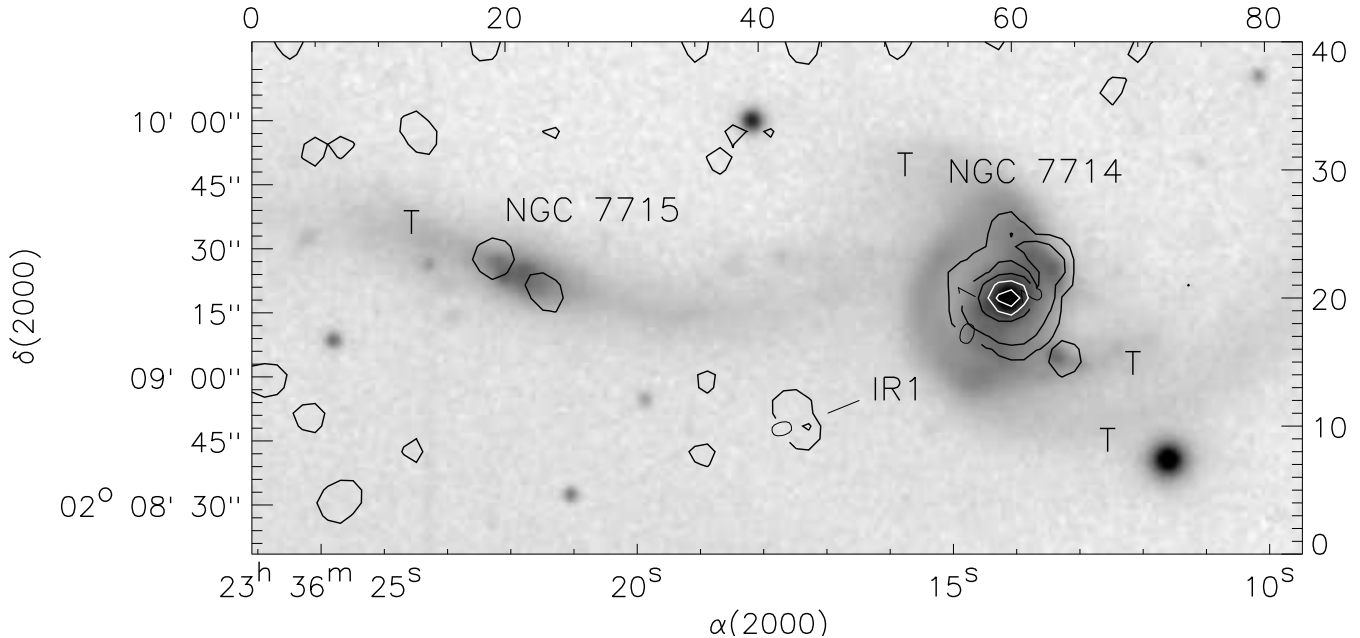}}
\caption{Deconvolved 15 $\mu$m LW3 map of \object{NGC 7714} and \object{NGC
7715} overlaid on an R band CCD image, showing the bridge between the two
galaxies.  Features denoted by 'T' give the positions of tails/arms eminating
from both galaxies.  Sources coinciding with \object{NGC 7715} were determined
to be glitches using robust deglitching techniques.  An enlarged view of the
\object{NGC 7714} region is given in Figure 2.  IR1 is a weak source without an
identified optical counterpart.}  \end{figure*}

The interacting system \object{Arp 284} consists of an active starburst galaxy
\object{NGC 7714} and its post starburst companion \object{NGC 7715}.  This
system has been the subject of several investigations
\cite{iso-demoulin-1968apj,iso-weedman-1981apj} because of its unusual
morphology.  Van Breugel et al. (1985) first reported weak Wolf-Rayet features
near 4846 \AA, and possible HeII emission from the nucleus of \object{NGC 7714}.
\object{NGC 7714} is an SBb peculiar galaxy and classified by Weedman et al.
(1981) as a proto-typical starburst.  The heliocentric radial velocity is 2798
km s$^{-1}$ which places it at a distance of 37.3 Mpc assuming H$_{o}$ = 75 km
s$^{-1}$ Mpc$^{-1}$ \cite{iso-gonzalez-1994apj}.  The spectrum from X-rays
\cite{iso-stevens-1998mnras} to VLA radio \cite{iso-smith-1992apj} at 6 and 20
cm, was explained as a result of intense star formation in the nucleus
\cite{iso-weedman-1981apj}.  Very detailed studies have been carried out in the
ultraviolet, optical and near infrared to quantify the gas properties in the
nuclear and circumnuclear regions
\cite{iso-weedman-1981apj,iso-gonzalez-1994apj,iso-garcia-1997apj}.  Hubble
Space Telescope (HST) spectroscopy of the nuclear starburst revealed Wolf-Rayet
features in the ultraviolet and indicate a population of about 2000 Wolf-Rayet
stars \cite{iso-garcia-1997apj}.  The B-magnitude of the galaxy is -20.04 and
far-infrared luminosity of 2.8 $\times$ 10$^{10}$ L$_{\odot}$ with IRAS
far-infrared flux ratios f$_{60}$/f$_{100}$ and f$_{25}$/f$_{60}$ of 0.9 and
0.25 respectively.  These ratios indicate that \object{NGC 7714}/7715 almost
qualifies as a 60 $\mu$m peaker source
\cite{iso-heisler-1996mnras,iso-laureijs-2000aa}.

 We present Infrared Space Observatory (ISO) observations of the Arp 284 system
containing the interacting galaxies \object{NGC 7714}/7715 as part of a program
investigating several galaxies exhibiting Wolf-Rayet signatures.  The
observations and data reduction are presented in Sect.  2.  The results
are contained in Sect.  3 and discussed in Sect.  4.  The
conclusions are summarised in Sect.  5.

\section{Observations and Data Reduction}

The ISO observations were obtained using the mid-infrared camera ISOCAM
\cite{iso-cesarsky-1996aa} and the spectrometric mode of the ISO
photopolarimeter ISOPHOT \cite{iso-lemke-1996aa}.  The astronomical observing
template (AOT) used was CAM 01, for the raster observations and PHOT40, for
spectrometry.  The log of the CAM01 observations using the LW3, LW6 and LW9
filters and the PHOT40 observations are presented in Table 1.  

\begin{table*} \caption{Log of the ISO observations of \object{NGC 7714}
and \object{NGC 7715}.  The nine columns list the observation name, the AOT, the
filter, the wavelength range ($\Delta\lambda$), the reference wavelength of the
filter, the date and duration of the observation, and the position of the
observation in RA and declination respectively.}

\begin{flushleft}
\begin{tabular}{lllllllll} \hline \noalign{\smallskip} Observation & AOT\# &
filter & $\Delta\lambda$ & $\lambda_\mathrm{ref}$ & date & duration & Right
Ascension & Declination \\ & & &($\mu$m) & ($\mu$m) & & (seconds)& (RA) &
(Dec)\\ \noalign{\smallskip} \hline \noalign{\smallskip} NGC7714 & CAM01 & LW3 &
12--18 & 14.3 & 24 May 1997 & 1936 & 23h 36m 18s & +02$^{\circ}$ 09$'$ 18$''$\\
NGC7714 & CAM01 & LW6 & 7--8.5 & 7.7 & 24 May 1997 & 1938 & 23h 36m 18s &
+02$^{\circ}$ 09$'$ 18$''$\\ NGC7714 & CAM01 & LW9 & 14--16 & 15 & 27 May 1997 &
1936 & 23h 36m 18s & +02$^{\circ}$ 09$'$ 18$''$\\ NGC7714 & PHT40 & SS/SL &
2.5--11.6 & & 24 May 1997 & 1132 & 23h 36m 14s & +02$^{\circ}$ 09$'$ 18$''$\\
NGC7715 & PHT40 & SS/SL & 2.5--11.6 & & 24 May 1997 & 1132 & 23h 36m 22s &
+02$^{\circ}$ 09$'$ 25$''$\\ \noalign{\smallskip}\hline \end{tabular}
\end{flushleft} \label{iso-log} \end{table*}

\subsection{ISOCAM}

Each observation had the following configuration:  3$''$ pixel field of
view (PFOV), integration time of 5.04 s with a $5''\times2''$ raster, 48$''$
stepsize, excluding discarded stabilization readouts
\cite{iso-siebenmorgen-1999esa}.  All data processing was performed with the CAM
Interactive Analysis (CIA) system \cite{iso-ott-1997asp,iso-delaney-1998esa},
and the following method was applied during data processing.  (i) Dark
subtraction was performed using a dark model with correction for slow drift of
the dark current throughout the mission.  (ii) Glitch effects due to cosmic rays
were removed following the method of Aussel et al.  (1999).  (iii) Transient
correction for flux attenuation due to the lag in the detector response was
performed by the method described by Abergel et al.  (1996).  (iv) The raster
map was flat-fielded using a library flat-field.  (v) Pixels affected by glitch
residuals and other persistent effects were manually suppressed.  (vi) The
raster mosaic images were deconvolved with a multi-resolution transform method
described by Starck et al.  (1998).  The duration of the observations in Table 1
give the length of the actual observation including instrumental, but not
spacecraft, overheads.  Photometry was performed by integrating the pixels
containing source flux exceeding the background by 3$\sigma$.  Contour levels
are based on a power scale, where the lowest level is approximately 2 times the
standard deviation, or $\sigma$, of the background noise, using pixels from
around the border of the image.

\subsection{PHOT-S}

 PHOT-S consists of a dual grating spectrometer with a resolving power of 90.
Band SS covers the range 2.5 - 4.8 $\mu$m, while band SL covers the range 5.8 -
11.6 $\mu$m.  \cite{iso-laureijs-1998esa}.  The PHT-S spectra of \object{NGC
7714} was obtained by pointing the $24''\times24''$ aperture of PHT-S
alternatively towards the peak of the LW6 emission (for 512 seconds) and then
towards two background positions off the galaxy (256 seconds each), using the
ISOPHOT focal plane chopper.  The calibration of the spectrum was performed by
using a spectral response function derived from several calibration stars of
different brightness observed in chopper mode
\cite{iso-acostapulido-1999kluwer}.  The relative spectrometric uncertainty of
the PHOT-S spectrum is about 20\% when comparing different parts of the spectrum
that are more than a few microns apart.  The absolute photometric uncertainty is
about 30\% for bright calibration sources.  All data processing was performed
using the ISOPHOT Interactive Analysis system, version 8.1
\cite{iso-gabriel-1998esa}.  Data reduction consisted primarily of the removal
of instrumental effects.  Once the instrumental effects have been removed,
background subtraction was performed and flux densities were obtained.  These
fluxes were plotted to obtain the spectrum for \object{NGC 7714}.

\section{Results}

\begin{figure}\resizebox{\columnwidth}{!}
{\includegraphics*{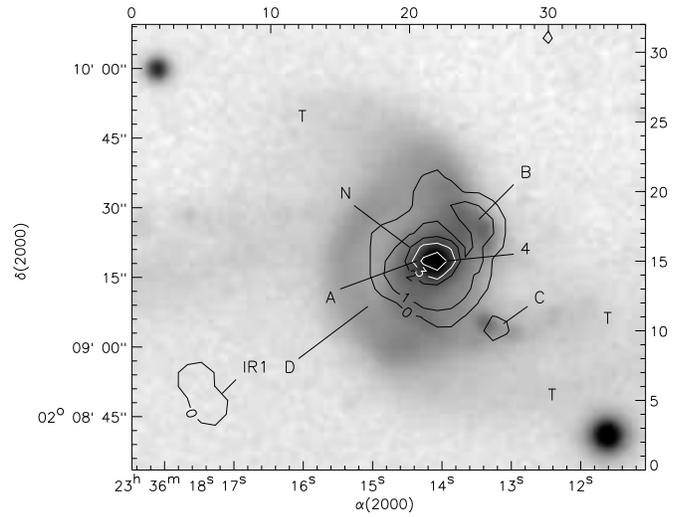}}
\caption{Deconvolved LW3 map of \object{NGC 7714}, superimposed on a R band CCD
image.  The three giant HII regions are marked A, B and C. D is the partial ring
and N denotes the nucleus with the Wolf-Rayet signature.  The contour levels
(mJy/arcsec\(^{-2}\)) are:  0 = 0.1, 1 = 1.0, 2 = 7.0, 3 = 20.0, 4 = 40.0.  IR1
is a weak source without an identified optical counterpart.}
\end{figure}

\begin{figure} \resizebox{\columnwidth}{!}
{\includegraphics*{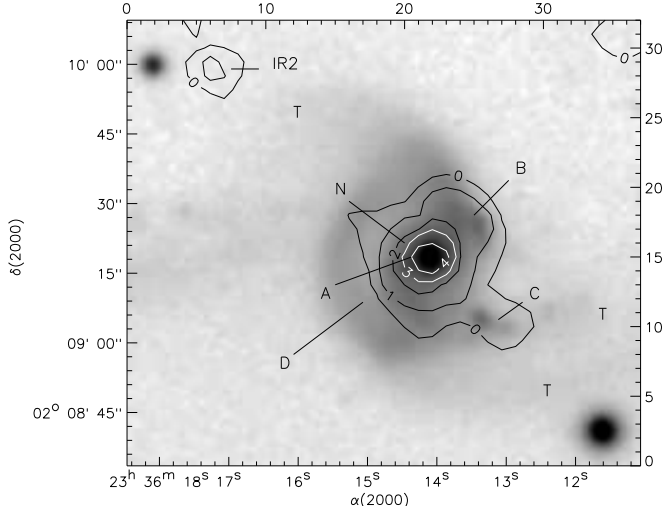}}
\caption{Deconvolved LW6 map of \object{NGC 7714}, superimposed on a R band CCD
image of \object{NGC 7714}.  Contour levels (mJy/arcsec\(^{-2}\)) are:  0 = 0.2,
1 = 1.0, 2 = 3.5, 3 = 9.0, 4 = 20.0.  The map notation is the same as in Fig.
2.  IR2 coincides with a weak radio source without an identified optical
counterpart.}  \end{figure}

\begin{table} \caption[]{ISOCAM and IRAS fluxes for \object{NGC 7714}.  The
four columns list the filter used, the wavelength range, the reference
wavelength for the filter and the flux measured.  The fluxes have a photometric
accuracy of about 15\%.  IRAS fluxes for \object{NGC 7714} are also given for
comparison \cite{iras-moshir-1990ipac}.}
\begin{flushleft} \begin{tabular}{llll} \hline\noalign{\smallskip} filter &
$\lambda$ & $\lambda_\mathrm{ref}$ & flux \\ & ($\mu$m) & ($\mu$m) & (Jy)  \\
\noalign{\smallskip}\hline\noalign{\smallskip} 
LW3 & 12.0--18.0 & 14.3 & 0.46 $\pm$ 0.07 \\
LW6 & 7.0--8.5 & 7.7 & 0.35 $\pm$ 0.05 \\ 
LW9 & 14--16 & 15 & 0.55 $\pm$ 0.08 \\
IRAS 12 $\mu$m & 8.5--15 & 12 & 0.47 $\pm$ 0.05 \\
IRAS 25 $\mu$m & 19--30 & 25 & 2.85 $\pm$ 0.26 \\
IRAS 60 $\mu$m & 40--80 & 60 & 10.36 $\pm$ 1.24 \\
IRAS 100 $\mu$m & 83--120 & 100 & 11.51 $\pm$ 0.69 \\
\noalign{\smallskip}\hline \end{tabular} \end{flushleft}
\label{isocam-fluxes} \end{table}

A deconvolved LW3 map overlaid on a R band CCD image \cite{iso-papaderos-1998aa}
of \object{Arp 284} is presented in Fig.1.  Deconvolved maps obtained using the
LW3, LW6 and LW9 filters, overlaid on a R band CCD image
\cite{iso-papaderos-1998aa}, are presented in Figs.  2, 3 and 4.  It is
evident that the strong compact source on each map coincides with the nuclear
region of \object{NGC 7714}.  Three giant H II regions, labelled A, B and C were
detected in all three maps, with A lying close to the nucleus, and B and C to
the northwest and southwest of the nuclear region \cite{iso-garcia-1997apj}.  It
is interesting to note that apart from a slight spur of LW6 emission, the bright
optical ring to the east of the nucleus was not detected, with infrared emisson
ceasing abruptly at the interface between the disk and the ring.  The spiral
arms of \object{NGC 7714} and the bridge linking the two galaxies were not
detected.  On the LW3 map, the source IR1 does not coincide with any known
optical counterpart, while the source IR2 on the LW6 map coincides with the
radio source RGB J2336+021 \cite{iso-laurent-1997aas}.

\begin{figure} \resizebox{\columnwidth}{!}
{\includegraphics*{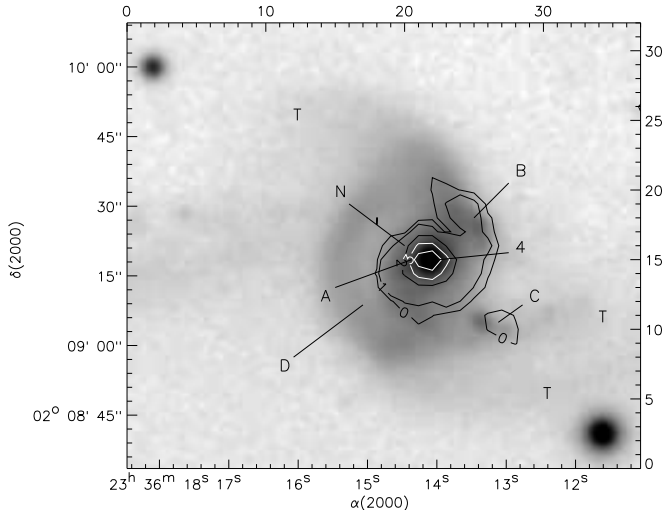}} \caption{LW9 map
of \object{NGC 7714}.  Contour levels (mJy/arcsec\(^{-2}\)) are:  0 = 0.4, 1 =
1.0, 2 = 12.0, 3 = 33.0, 4 = 70.0.  The map notation is the same as in Fig. 2.
The nuclear region coincides with the strong LW9 emission.}  \end{figure}

The companion galaxy \object{NGC 7715} was not detected in any of the three
filter bands.  Two weak sources were detected at the position of \object{NGC
7715} in the LW3 filter in Fig. 1, but were determined to be glitches using
robust deglitching techniques.  The lack of emission from \object{NGC 7715},
along with the ring and spiral arms of \object{NGC 7714} suggests that strong
star formation is not present in these regions.  

The PHOT-SL spectrum of \object{NGC 7714} is presented in Fig.  5.  Strong
detections were made of unidentified infrared bands (UIBs) at 6.2, 7.7, 8.6 and
11.3 $\mu$m that are usually attributed to polycyclic aromatic hydrocarbons
(PAHs) \cite{iso-allamandola-1989apjs}.  A weak additional feature at 11.0
$\mu$m may also be PAH \cite{iso-moutou-1999edp}.  The 11.3 $\mu$m feature is
quite strong relative to the 7.7 $\mu$m, with a flux ratio of 2:1.  [Ar II] was
detected at 6.99 $\mu$m in the unsmoothed spectrum at the 3 $\sigma$ level.  Two
weak features were also detected at about 9.66 and 10.6 $\mu$m.  The 9.66 $\mu$m
feature may be due to the S(3) pure rotational line, $\upsilon$ = 0-0, of
molecular hydrogen, while the 10.6 $\mu$m feature may be a blend of [S IV] at
10.51 $\mu$m and an additional unidentified component at 10.6 $\mu$m to account
for the width of the feature.  The interpolated continuum between the ends of
the wavelength range was subtracted to obtain the line fluxes.  The identified
line features and line fluxes are given in Table 3.

\begin{figure} \resizebox{\columnwidth}{!}
{\includegraphics*{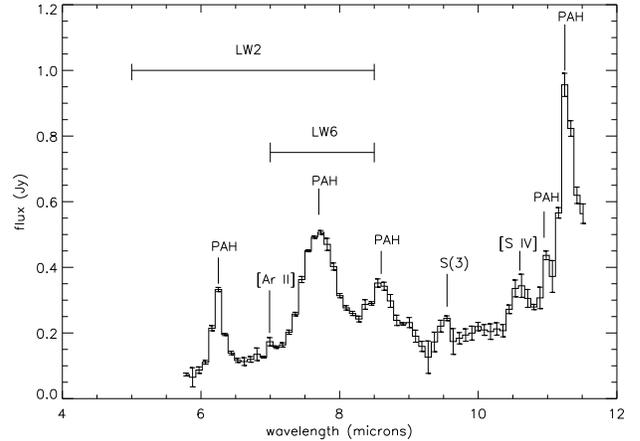}}
\caption{Unsmooothed PHOT-SL spectrum of \object{NGC 7714}.  The PAH
features, along with [Ar II], [S IV] and the S(3) ground vibrational state of
molecular hydrogen at 9.66 $\mu$m are indicated.  The bandpasses for the LW2 and
LW6 ISOCAM filters are also indicated.}  \end{figure}

\begin{table} \caption[]{PHOT-S fluxes for \object{NGC 7714}. The three 
columns give the line identification, the wavelength range and the integrated 
flux respectively.}

\begin{flushleft} \begin{tabular}{lll} \hline\noalign{\smallskip} line ID &
Wavelength range& flux  \\ & ($\mu$m) & (10$^{-15}$ W m$^{-2}$) \\
\noalign{\smallskip} \hline\noalign{\smallskip} 
PAH 6.2 $\mu$m & 6.0 - 6.6 & 2.73 $\pm$ 0.46 \\ 
$[$Ar II$]$ 6.99 $\mu$m & 6.8 - 7.2 & 0.74 $\pm$ 0.10 \\
PAH 7.7 $\mu$m & 7.2 - 8.3 & 8.84 $\pm$ 1.10 \\ 
PAH 8.6 $\mu$m & 8.3 - 8.9 & 1.22 $\pm$ 0.24 \\
H$_{2}$ 9.66 $\mu$m & 9.3 - 9.9 & 0.78 $\pm$ 0.20 \\ 
$[$S IV$]$ 10.5 $\mu$m & 10.4 - 10.8 & 0.32 $\pm$ 0.08 \\ 
PAH 11.0 $\mu$m & 10.8 - 11.1 & 0.23 $\pm$ 0.07 \\ 
PAH 11.3 $\mu$m & 11.1 - 11.6 & 1.28 $\pm$ 0.26 \\ 
\noalign{\smallskip}\hline \end{tabular} \end{flushleft}
\label{isophot-fluxes} \end{table}

The spectral energy distribution of \object{NGC 7714} using ISOCAM, PHT-SL and
IRAS fluxes is presented in Fig. 6.  The dust model, denoted by the solid
curve \cite{iso-krugel-1994aa,iso-siebenmorgen-1998aa}, contains three separate
dust populations:  the PAH bands \cite{iso-boulanger-1998aa}, a warm dust
component at 110 K and a cooler dust component at 40 K and each component is
indicated in the figure.

\begin{figure} \resizebox{\columnwidth}{!}
{\includegraphics*{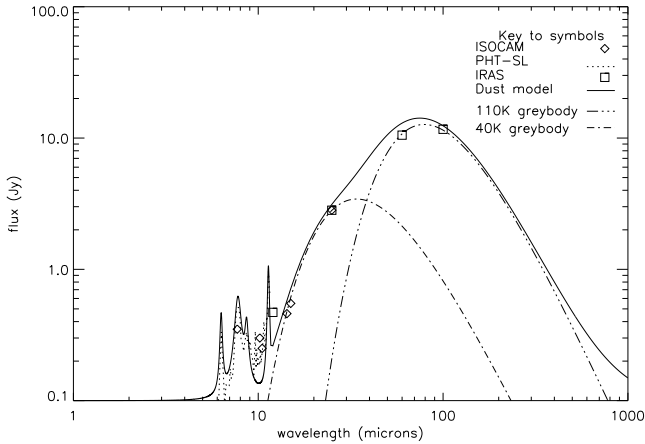}}
\caption{Spectral energy distribution of \object{NGC 7714} using ISO and IRAS
flux data, including the key for the different symbols.  The models are
described in the text.}  \end{figure}

\section{Discussion}

In order to explain the observed morphology and kinematics, Smith \& Wallin
(1992) modelled the \object{Arp 284} system using an off-centre parabolic
collision between two disk galaxies with a mass ratio \(M_{2}/M_{1}\) $\sim$
0.3.  The interaction occured $\sim$ 1.1 x \(10^{8}\) years ago, with the
intruder galaxy impacting at a distance of 0.85 times the radius of the target
disk.  This model successfully reproduced the observed morphology, including the
partial ring and the bridge linking the two galaxies.  Further modelling
\cite{iso-smith-1997apj} showed the bridge to be a hybrid between the tidal
bridges observed in systems such as M 51 and gaseous 'splash' bridges such as
that in the 'Taffy' interacting system VV 254 \cite{iso-jarrett-1999esa}.  The
'Taffy' system, which is at a comparable distance of 58.4 Mpc, is intrinsically
bright in the mid-infrared and consists of three principal morphological
infrared components:  two multipeaked nuclear regions, a large scale ring or
wrapped spiral arm and a bridge linking the two galaxies which contains one
quarter of the total H I in the system and has a significant dust content
\cite{iso-jarrett-1999esa}.  Unlike the 'Taffy' system however, the bridge
linking \object{NGC 7714} and \object{NGC 7715} was not detected in any of our
three ISOCAM bands.  Both \object{Arp 284} and the 'Taffy' system have
comparable HI column densities \cite{iso-smith-1997apj}, but the lack of
detectable emission from the bridge of \object{Arp 284} indicates a low warm 
dust content. Further observations at longer wavelengths are required to 
determine the extent of the cold dust component.

The partial ring within NGC 7714, populated by red giant stars like those found
in the companion galaxy, seems to be deficient in warm dust and is not
detectable in the CAM maps.  The model of Smith \& Wallin (1992) may explain
the lack of ongoing star formation within the partial ring of \object{NGC 7714}.
In more central impacts, occuring at less than 0.2 times the radius of the
target disk, the resulting strong radial oscillations lead to the formation of
rings similar to those observed in the \object{Cartwheel} and \object{Arp 10}
\cite{iso-charmandaris-1999esa}.  Smith et al. (1997) noted that the
optical ring in NGC 7714 does not have a prominent H I counterpart, similar to
that found in the non star-forming inner ring of the Cartwheel but not in the
outer ring \cite{iso-higdon-1996apj}.  Such differentiation of gas and stars is
expected in partial rings formed during very off-center collisions.  Thus, the
lack of an H I counterpart does not rule out a collisional origin for the NGC
7714 ring.  Another possibility is that the stellar ring may be a wrapped-around
spiral arm caused by a noncollisional prograde planar encounter, rather than a
collisional ring.  The high column density of gas in the bridge, however, and
its offset from the stars \cite{iso-smith-1997apj}, supports the concept that
gas was forced out of the main galaxy by a collision between two gas disks
rather than merely perturbed in a grazing or long-range encounter.

The age of Wolf-Rayet stars in the central burst of \object{NGC 7714} has been
dated at between 4 and 5 Myr \cite{iso-garcia-1997apj}, yet this age is far
short of the $\sim$ 10$^{8}$ years since the interaction.  Given the amount of
time since the onset of the interaction, any burst of star formation initiated
by the interaction should have long since ceased, for example the companion
galaxy \object{NGC 7715}, which has been in a post starburst phase for the last
50 - 70 Myr \cite{iso-bernlohr-1993aa}.  For such a young burst, gas infall to
the nucleus must be ongoing to power the burst, and indeed HI gas may be falling
from the bridge to the nuclear region
\cite{iso-papaderos-1998aa,iso-smith-1992apj}.  In a survey of 10 interacting
galaxies, Bushouse et al.  (1998) noted that galaxies similar in morphological
type to \object{NGC 7714} possessed nuclear infrared sources that are 10-100
times brighter than normal galaxies in the mid-infrared and have high levels of
star formation.  The companion galaxies, such as \object{NGC 7715}, have passed
through their own star formation epoch and now lack the gas and dust to suport
large-scale star formation.

\begin{figure} \resizebox{\columnwidth}{!}
{\includegraphics*{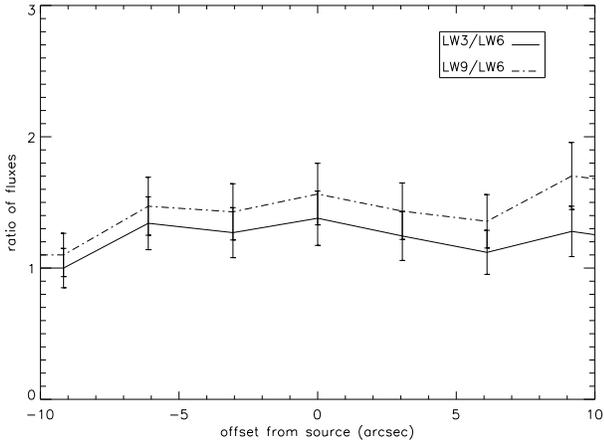}} \caption{Ratios
of ISOCAM emission profiles in three filters obtained from the NE to SW scan
across \object{NGC 7714} at PA = 45$^{\circ}$.  The solid line gives the
LW3/LW6 ratio, while the dot-dash line gives the LW9/LW6 ratio, and the
1$\sigma$ error bars are included.}  \end{figure}

It is interesting to note that the ratio of LW3/LW2, where LW3 emission is
dominated by dust and LW2 by PAHs, varies depending on the separation between
the interacting galaxies \cite{iso-hwang-1999apj}.  The LW3/LW2 ratio generally
decreases as interactions develop, starbursts age and separations increase
\cite{iso-vigroux-1999esa,iso-cesarsky-1999kluwer,iso-charmandaris-1999asp}.  To
determine the LW3/LW2 ratio for \object{NGC 7714}, the LW2 flux was synthesized
from the PHOT-SL spectrum in the equivalent range of wavelengths to the ISOCAM
LW2 filter, 5 to 8.5 $\mu$m.  In order to check the accuracy of the LW2 flux,
LW2 and LW6 fluxes were obtained for several sources, including \object{Mkn 297}
and \object{NGC 1741}, from their PHOT-S spectra and the derived values agree
with the respective ISOCAM fluxes within the photometric uncertainties.  For
\object{NGC 7714}, the LW3/LW2 ratio was 1.5$\pm$0.2.  This is quite low, since
a ratio above 3 is indicative of high star formation due to the heating of dust
in the nuclear region by hot, young ionising stars \cite{iso-vigroux-1999esa}.

Since the LW3 and LW9 fluxes are dominated by dust emission, and the LW2 and LW6
by PAHs, the LW3/LW6 and LW9/LW6 ratios provide diagnostics similar to the
LW3/LW2 ratio, allowing a determination of the current state of the starburst.
To determine the LW9/LW6 and LW3/LW6 ratios, pixel fluxes were obtained in
strips on each of the three raster maps, each strip bisecting \object{NGC 7714}
through the nuclear region.  Each strip was equally separated by a position
angle (PA) of 45$^{\circ}$, measured from north, where PA = 0$^{\circ}$.  The
strips were then divided to obtain LW9/LW6 and LW3/LW6 ratios.  One such strip
is presented in Fig. 7, taken from northwest to southeast through the galaxy.
The LW9/LW6 and LW3/LW6 ratios were quite low, with values falling between 1 and
1.7.  Combining these findings with the low value of 1.5$\pm$0.2 for the LW3/LW2
ratio, it may be that while the nuclear region of \object{NGC 7714} is still in
the throes of a young starburst indicated by the Wolf-Rayet features, the low
ratio values along with the age of the burst suggest that
\object{NGC 7714} is moving into a post-starburst stage
\cite{iso-charmandaris-1999asp}.  The aging of the burst may be due to
supernovae and stellar winds \cite{iso-taniguichi-1988aj} disrupting the
interstellar medium, preventing further star formation as soon as the first
generation of O stars have formed and evolved \cite{iso-genzel-2000astroph}.

The 11.3 $\mu$m PAH feature is quite strong relative to the 7.7 $\mu$m PAH
feature.  This ratio is particularly sensitive to the degree of ionization,
implying that the detected PAHs may not have been ionized.  A strong 11.3 $\mu$m
feature is common in colder galaxies like \object{NGC 7714}, since it is linked
to neutral PAHs and is indicative of a high degree of hydrogenation in the PAHs
\cite{iso-peeters-1999esa,iso-lu-1999esa}.  PAHs exposed to a harder radiation
field, for example within the H II regions and on the interface between the H II
and the molecular cloud, can be ionized, loose hydrogen atoms and/or disappear
by photodissociation.  Proximity to highly ionizing sources, such as young O
stars within a burst may thus lead to dehydrogenation of the PAHs.  Moving out
past the H II/molecular cloud interface, the degree of hydrogenation increases
\cite{iso-verstraete-1996aa,iso-roelfesma-1996aa}.

The compact nature of the nucleus of \object{NGC 7714} and its extremely high
surface brightness \cite{iso-weedman-1998aj} have led to suspicions that
\object{NGC 7714} harbours an active nucleus.  While the galaxy has far-infrared
IRAS colours more in line with Seyfert 2 galaxies than starbursts
\cite{iso-gonzalez-1994apj}, the emission lines of H$\alpha$, [NII]
$\lambda$$\lambda$6548, 6583, [SII] $\lambda$$\lambda$6716, 6731, He I
$\lambda$5876 and [OI] $\lambda$ 6300 are not broad enough to suggest the
presence of a Seyfert nucleus \cite{iso-demoulin-1968apj}.  UV observations by
the HST \cite{iso-gonzalez-1999apj} have shown that the nuclear region contains
$\sim$ 2000 Wolf-Rayet and 20000 O type stars, with an age for the burst of 4 to
5 Myr, while ROSAT \cite{iso-papaderos-1998aa,iso-stevens-1998mnras} has shown
it to be a strong X-ray source.  The high excitation of material within the
nuclear region can account for the 6.99 $\mu$m [Ar II] line.  The 9.6 $\mu$m
feature may be due to the S(3) ground vibrational molecular hydrogen ground
state of molecular hydrogen.  Ultraviolet fluorescence, excitation by
low-velocity shocks and heating caused by X-rays are considered to be the
primary emission mechanisms for the excitation of molecular hydrogen
\cite{iso-black-1987apj,iso-draine-1982apj,iso-mouri-1994apj}.  These mechanisms
require a source of dense gas to be located near the source of illumination such
as a starburst.  Excitation may also occur due to slow shocks induced by jets or
by kinetic processes such as winds, superwinds and supernovae.  Spectroscopic
studies of Seyfert galaxies indicate that no single process in responsible for
the H$_{2}$ emission \cite{iso-quillen-1999apj}.

Several diagnostic tools are available to probe the nature of the activity
within the nuclear region.  The ratio of the integrated PAH luminosity and the
40 to 120 $\mu$m IR luminosity \cite{iso-lu-1999esa} provides a tool to
discriminate between starbursts, AGN and normal galaxies because the lower the
ratio, the more active the galaxy.  For \object{NGC 7714}, the ratio is 0.09 and
is consistent with the values found for other starbursts
\cite{iso-vigroux-1999esa}.  Similarly, the ratio of the 7.7 $\mu$m PAH flux to
the continuum level at this wavelength can provide a measure of the level of
activity within the nucleus \cite{iso-genzel-1998apj,iso-laureijs-2000aa}.  The
ratio for \object{NGC 7714} is 3.3 and is indicative of an active starburst.
The results from these diagnostic tools indicate that \object{NGC 7714} is home
to a compact burst of star formation, and suggests that an AGN is not present
within the nuclear region.

\section{Conclusions}

 Deconvolved ISOCAM maps of the \object{Arp 284} system were obtained with
the LW3, LW6 and LW9 filters, with strong emission detected from the central
source in \object{NGC 7714}.  IR emission was not detected from the companion
galaxy \object{NGC 7715}, the bridge linking the two galaxies or from the
partial stellar ring, where emission ceases abruptly at the interface between
the disk and the ring.  ISOPHOT spectrometry of the nuclear region of
\object{NGC 7714} was also obtained, with strong PAH features, the emission line
[Ar II], molecular hydrogen at 9.66 $\mu$m and a blend of lines including [S IV]
about 10.6 $\mu$m all present within the spectrum.

The morphology of the system is well described by an off-centre collision
between the two galaxies.  A series of diagnostic tools allowed an investigation
to be performed regarding the activity within the central region of \object{NGC
7714}.  The LW3/LW2, where the LW2 flux was synthesized from PHOT-S
measurements, LW9/LW6 and LW3/LW6 ratios suggest that the central burst within
\object{NGC 7714} is moving towards the post-starburst phase.  The ratio of the
integrated PAH luminosity to the 40 to 120 $\mu$m infrared liminosity and the
far-infrared colours reveal that despite the high surface brightness of the
nucleus, the properties of \object{NGC 7714} can be explained in terms of a
starburst and do not require the prescence an AGN.

\begin{acknowledgements}

We thank P.  Papaderos and K.  Fricke for kindly allowing the use of their CCD
image of the \object{Arp 284} system.  The ISOCAM data presented in this paper
was analysed using `CIA', a joint development by the ESA Astrophysics Division
and the ISOCAM Consortium.  The ISOCAM Consortium is led by the ISOCAM PI, C.
Cesarsky, Direction des Sciences de la Matiere, C.E.A., France.  The ISOPHOT
data was reduced using PIA, which is a joint development by the ESA Astrophysics
Division and the ISOPHOT consortium.

\end{acknowledgements}


\end{document}